\def \m2   {\mu^{2 \epsilon}}
\def\lQ{\Lambda_{\rm QCD}}
\def\als{\alpha_{\rm s}} 
\def\bea{\begin{eqnarray}}
\def\eea{\end{eqnarray}}
\def\be{\begin{equation}}
\def\ee{\end{equation}}
   
\def\la{\lambda} \def\lap{\lambda^{\prime}}  \def\de{\delta}  \def\dag{\dagger}
    
\def\bnabla{{\bm \nabla}}

\documentclass[a4paper, 11pt]{article}
\usepackage{pos}

\usepackage{bm}
\usepackage{bbm}
\usepackage{braket}
\usepackage{slashed}

\title{Effective Field Theories and Lattice QCD for the X Y Z frontier}

\author*[a,b,c]   {Nora Brambilla}

\affiliation[a]{Physik Department, Technische Universit\"at M\"unchen,\\
James-Franck-Strasse 1, 85748 Garching, Germany}

\affiliation[b]{Institute for Advanced Study, Technische Universit\"at M\"unchen,\\
Lichtenbergstrasse 2 a, 85748 Garching, Germany}
  
\affiliation[c]{Munich Data Science Institute, Technische Universit\"at M\"unchen, \\  
Walther-von-Dyck-Strasse 10, 85748 Garching, Germany}

\emailAdd{nora.brambilla@ph.tum.de}

\abstract{Exotic states have been predicted before and after the advent of QCD.
In the last decades they have been observed at accelerator experiments  in
the sector with two heavy quarks, at or above the quarkonium strong decay
threshold and called X Y Z states.
These states  offer a unique possibility for investigating the dynamical
properties of strongly correlated systems in QCD.
I will show how an alliance of nonrelativistic effective field theories and
lattice can allow us to address these states in QCD. In particular I will
explain what are  the opportunities and challenges of lattice QCD in this
respect and which new tools should be  developed.}

\FullConference{%
 The 38th International Symposium on Lattice Field Theory, LATTICE2021
  26th-30th July, 2021
  Zoom/Gather@Massachusetts Institute of Technology
}

\begin{document}
\maketitle

\section{Introduction}
Exotic states, i.e. states   with a composition different 
from a quark-antiquark or three quarks in  a color singlet,
have been predicted before and after the inception of QCD.
In  the last decades  a large number of states, either with a manifest different composition (with isospin
and electric charge  different from zero) or with other exotic characteristics 
have been observed  in the sector
with two heavy quarks $Q \bar{Q}$, at or above the quarkonium strong decay threshold at the
B-Factories, tau-charm and LHC  and Tevatron collider experiments \cite{Brambilla:2019esw,Brambilla:2010cs,QuarkoniumWorkingGroup:2004kpm,Brambilla:2014jmp}.
These states have been termed $X, Y, Z$  in the  discovery publications,
without any special criterion, apart 
from  $Y$  being used for exotics with vector
 quantum numbers, i.e., $J^{PC} = 1^{--}$ (where $J$ is the spin,
 $P$ is the parity and $C$ is the charge-conjugation quantum
 number). Meanwhile, the Particle Data Group (PDG) has proposed a new naming scheme
 \cite{ParticleDataGroup:2018ovx},  that extends the scheme used for ordinary quarkonia, in which
 the new names carry information on the $J^{PC}$ quantum numbers, see \cite{Brambilla:2019esw} for more details.

Some of these exotics have quantum numbers that cannot be obtained with ordinary hadrons. In this case, the
identification of these states as exotic is straightforward. In the other cases, the distinction requires a careful analysis
of experimental observations and theoretical predictions.
Of course all hadrons should be color singlets but allowing combinations beyond the $Q \bar{Q}$ and $QQQ$ in the two heavy quarks sector
calls for  tetraquarks like  $Q \bar{Q}  q \bar{q}$, $QQ  \bar{q} \bar{q}$,  pentaquarks  like $Q \bar{Q}  q q q$, hybrids $Q \bar{Q}  g$
and so on \cite{Ali:2017jda} .  We have observed these exotics up to now  only in the sector with two heavy quarks likely due to the fact 
that the presence of the two heavy quarks stabilizes them.  
{\it Some of the discovered states have an  unprecedentely  and surprinsingly small width even if they are at  or above the strong decay threshold.
$ X Y Z$ states  offer us unique possibilities  for the investigation of  the dynamical properties of
strongly correlated systems in QCD:  we have by now measured  their signatures  in the spectrum and we should 
develop the tools to gain a solid interpretation from the underlying field theory, QCD.
This is a very significant problem with trade off to other fields featuring strong correlations and a pretty interesting 
connections to  heavy ion physics, as propagation of these states in medium may help us to scrutinize their properties.}

Since  the new quarkonium revolution i.e. the discovery  of the first exotic state, the $X(3872)$  at BELLE in 2003 \cite{Belle:2003nnu},
a wealth of theoretical papers appeared  to  supply interpretation  and understanding of  the characteristics of the exotics.
Many models  are based  on the choice of  some dominant degrees  of freedom and an assumption on   the related interaction 
hamiltonian. An effective field theory  molecular description of some of these states particularly close to threshold was also put forward, see e.g.
\cite{Guo:2017jvc,Braaten:2007dw,Braaten:2003he,Fleming:2011xa}. 
A priori the simplest system consisting of only two quarks and two antiquarks (generically called tetraquarks)
 is already a very complicated object and it is unclear whether or not any kind of clustering occurs in it. 
 However, to simplify the problem it is common to focus on certain substructures and investigate their implications:
 in hadroquarkonia the heavy quark and antiquark form a compact core surrounded by a light-quark cloud; in compact tetraquarks the relevant
 degrees of freedom are compact 
diquarks and antidiquarks; in the molecular picture two color singlet mesons are interacting at some typical distance.
Discussions about exotics therefore often concentrate on the  'pictures' of the states,
like for example the   tetraquark  interpretation against the molecular one (of which both several different realizations exist). 
However,  as a matter of fact all the light  degrees of freedom  (light quarks, glue, in different configurations)   should  be there in QCD 
close or above the strong decay threshold: it  is a   result  of the strong dynamics which one sets  in  and in which configuration   dominates  in a given regime.
Even in an ordinary quarkonium or in a heavy baryon, which has a dominant  $Q\bar{Q}$  or $QQQ$ configuration,  subleading contributions of the
quantum field theoretical Fock space may contribute,  with  have additional quark-antiquark pairs and active gluons.
{\it Addressed like that, the problem  turns out to be extremely challenging. 
I will show  how the presence of two heavy quarks
gives us a way to simplify the problem and to address it 
using nonrelativistic effective field theories (NR EFTs) and lattice.
The existence of a large scale, the scale of the mass of the heavy quark $m$, allows  us to take advantage of scale factorization and to use lattice QCD to evaluate purely glue dependent correlators at the 
low energy scale. Such correlators are not depending on flavor and are universal, i.e. the can be evaluated once and used for prediction on a large number of states.
 I will argue that this method is complementary and often preferable with respect to an ab initio lattice  evaluation of the properties of each single state.}
 To pursue this aim, i.e. to perform lattice evaluation directly  on factorized low energy objects in the framework of EFTs, we have founded a dedicated lattice collaboration called TUMQCD
 \cite{tumqcd}.
 
 The paper is organized in the following  way: in section two  I summarize  how the combination of NREFTs and lattice has allowed us to address
 quarkonium properties  in vacuum and in medium  in QCD.
In the following sections I put forward a similar framework for exotica.
In particular in section three I introduce  the so called BOEFT (Born Oppenheimer Effective Field Theory) 
to describe hybrids and I report the predictions of the theory and how  progress depends
on more precise and/or new lattice evaluations of low energy correlators, even quenched.
In Section four I extend this description to states with light quarks, which would give a theory of  the tetraquarks and molecular pictures.
In also examine how heavy ion collisions can help us supplying complementary information and what is the role of the EFT and lattice in this.
A list of lattice calculations that would be needed to gain progress is given in each section and in the outlook. In this latter
I also discuss tools, methods and open challenges to get these low energy correlators from the lattice.

\section{NREFTs and Lattice for Quarkonium} 

The study of quarkonium in the last few decades has witnessed two major developments: 
the establishment of NREFTs and  progress in lattice QCD calculations of excited states and resonances, 
with calculations at light-quark masses. 
Both allow for precise and systematically improvable computations that are (to a large extent) model-independent. 
It is  precisely this advancement in the understanding of quarkonium and quarkonium-like systems  inside QCD
that makes quarkonium exotics  particularly valuable.
{\it In fact, today that we are confronted with a huge amount of high-quality data, which
have provided for the first time uncontroversial evidence for the existence of exotic hadrons,
by  using modern theoretical tools that allow us to explore in a controlled way these new forms of matter we can get a
unique insight into the low-energy dynamics of QCD}.

\subsection{Physical scales of quarkonium}

Heavy quarkonia  are systems composed by two heavy quarks (charm, bottom), 
with mass $m$ larger than the ``QCD  confinement scale'' $\lQ$, 
so that $\als(m) \ll 1$ holds. From the quarkonia spectra it is evident that 
the difference in energy levels is much smaller than the quark mass and  therefore quarkonia are nonrelativistic 
systems. As such they are  multiscale systems.
 Being nonrelativistic, quarkonia are characterized by a small parameter, 
the heavy-quark relative velocity $v = |\vec{v}|$ in the rest frame of the meson, where 
 $v^2 \sim 0.1$ for the $b\bar{b}$, $v^2 \sim 0.3$ for $c\bar{c}$ systems. This parameter induces a 
hierarchy of dynamically generated energy scales: starting from the mass $m$ of the heavy quark (hard scale),
there is the  typical relative momentum $p \sim m v$, corresponding to the inverse Bohr radius $r \sim 1/(m v)$ (soft scale),
and a typical binding  energy $E \sim m v^2$ (ultrasoft (US)  scale). 
The description of the heavy $Q\bar{Q}$  systems depends finally on the relation of  $\lQ$ to the above mentioned 
scales. Clearly,  for energy scales close to $\lQ$ there is no longer any perturbative description  
and one has to rely on nonperturbative 
methods. Regardless of this, the nonrelativistic 
hierarchy $m \gg m v \gg m v^2$ 
persists also below the $\lQ$ threshold, as long as $v$ is small. 
While the hard scale   $m$ is always assumed to be  larger than 
$\lQ$, different  situations may arise for the other two scales.

The hierarchy of nonrelativistic scales makes the 
difference between heavy quarkonia and heavy-light mesons, which are characterized
by just two scales: $m$ and $\lQ$.
This renders the theoretical description of 
quarkonium  much more complicated.
All these scales get entangled in a typical 
amplitude involving a quarkonium
observable. Moreover   the nonrelativistic bound state problem is ``nonperturbative'',
even at small $\alpha_s$,  i.e.  infinite series of diagram needs to be resummed,
differently from an on shell scattering calculation.
Quarkonium annihilation 
and production take place  at the scale $m$,  
quarkonium binding takes place at the scale
$mv$, which is the typical momentum exchanged inside the bound state, while 
very low-energy (US) gluons and light quarks   
live long enough that a bound state has time to form and, 
therefore, are sensitive to the 
scale $mv^2$. US gluons are responsible for phenomena similar 
to the  the Lamb shift in QED. To address this nonrelativistic multiscale  system in
quantum  field theory is a  real challenge. It is so even in QED with positronium and
it is even more difficult here with the nonperturbative effects. Notice that even if you were
interested in a lattice first principle simulations, still the existence of a set of widely separated scales
will make the lattice calculation very  challenging,
The solution is to take advantage of the existence of the different  energy scales   
to substitute QCD with simpler but equivalent NREFTs.
A hierarchy of  NREFTs may be constructed by systematically integrating out 
modes associated with high-energy scales 
not relevant for the quarkonium system.
Such integration is made in a matching procedure that 
enforces the equivalence between QCD and the EFT at a given 
order of the expansion in $v$.
The EFT Lagrangian is factorized in matching coefficients,
encoding the high energy degrees of freedom  and low energy operators.
The relativistic invariance is realized via exact relations 
among the matching coefficients
~\cite{Manohar:1997qy,Brambilla:2003nt}.
The EFT displays a power counting in the small parameter $v$, therefore we are able 
to attach a definite power of $v$ to the contribution of each EFT operators to 
the physical observables.

\subsection{Physics at the scale $m$: NRQCD}
Nonrelativistic QCD (NRQCD)~\cite{Caswell:1985ui,Bodwin:1994jh}, 
follows from QCD  integrating out the scale $m$. As a consequence, 
the effective Lagrangian is organized as an 
expansion in $1/m$  and $\als(m)$: 

\begin{equation} \label{EQ:NRQCD}
{\cal L}_{\rm NRQCD}  = \sum_n \frac{c_n(\als(m),\mu)}{m^{n} } 
\times  O_n(\mu,mv,mv^2,...),
\end{equation}
where $c_n$ are Wilson coefficients that contain the 
contributions from the scale $m$; they  
can be perturbatively calculated by matching the full QCD result to 
the EFT. The  $O_n$ are  local operators of NRQCD; the matrix 
elements of these operators contain the
physics of scales below $m$, in particular of the scales  $mv$
and $mv^2$ and also of the nonperturbative scale $\lQ$. Finally, 
the parameter $\mu$ is the NRQCD factorization 
scale. The low-energy operators $O_n$  are constructed out of two or four  
heavy quark/antiquark fields plus gluons.
The quarkonium states $ \vert H\rangle$ in 
NRQCD is expanded in the number of partons
\begin{equation}
| H \rangle = | \bar{Q} Q  \rangle + | \bar{Q} Q  g \rangle + | \bar{Q} Q  \bar{q} q \rangle + \cdots  
\end{equation}  
where the states including one or more light parton are shown to be suppressed by powers of $v$.
In the  $| \bar{Q} Q  g \rangle$ for example the quark-antiquark are in a color octet state.
The NRQCD lagrangian has been extensively used on the lattice with great success  by the HPQCD collaboration 
to calculate quarkonium spectra and decays\footnote{Notice that
  the NRQCD leading piece of the Lagrangian differently from the HQET one is not renormalizable, so that lattice calculation cannot go to
  the continuum limit but have to stay in the scaling window.}.
On the other hand,  NRQCD has been deeply impactful on the study of quarkonium production at the LHC putting forward a factorization formula for
the inclusive cross section for the direct production of the
quarkonium $H$  at large momentum in the center of mass frame
 written as a sum of products of NRQCD matrix elements and 
short-distance coefficients:
\begin{equation} \label{EQ:Prod} 
\sigma[H]=\sum_n \sigma_n \langle {\cal K}_n^{4 {\rm fermions}} \rangle
\end{equation}
where  the
$\sigma_n$ are short-distance coefficients, and the matrix elements
$\langle {\cal K}_n^{4 {\rm fermions}} \rangle$ 
are vacuum-expectation values of objects similar to the four-fermion
operators in decays and containing both color singlet and color octet contributions.
The matrix elements (LDMEs, long distance matrix elements)  $\langle {\cal K}_n^{4 {\rm fermions}} \rangle $
contain all of the nonperturbative physics associated
with the  evolution of the $Q\bar Q$ pair into a quarkonium state. 
{\it Notice however that it has never been possible up to now  to calculate these LDMEs on the lattice 
due to their cumbersome definition.}

\subsection{Physics at the scale $mv$: pNRQCD}

Quarkonium formation happens at the scale $mv$. 
The suitable EFT is potential non relativistic QCD,  
pNRQCD \cite{Pineda:1997bj,Brambilla:1999xf,Brambilla:2004jw},  which 
follows from NRQCD by integrating out the scale $mv \sim r^{-1}$.
The soft scale $mv$  may be either larger or smaller than the confinement 
scale $\lQ$ depending on the radius of the quarkonium system. 
When $mv \gg \lQ$, we speak about weakly-coupled pNRQCD because 
the soft scale is perturbative and the matching from NRQCD to pNRQCD 
may be performed in perturbation theory. 
When $mv  \sim \lQ$, we speak about  
strongly-coupled pNRQCD because the soft scale 
is nonperturbative and the matching 
from NRQCD to pNRQCD is nonperturbative and cannot be calculated
with an expansion in $\alpha_s$.
{\it The  low energy nonperturbative factorized effects depend on the size of the physical system: 
when $\lQ$ is  comparable or smaller than $mv$ they are carried by correlators of 
chromoelectric  or chromomagnetic fields nonocal or local in time, when $\lQ$ is of order $mv$ they are carried 
by generalized static Wilson loops with insertion on chromoelectric and chromomagnetic fields.}
The EFT allows us to make model independent predictions and we can use the power counting to attach an error to the 
theoretical prediction. {\it The nonperturbative phyiscs in pNRQCD is encoded in few low energy correlators that depend only on the 
glue and are gauge invariant: these are objects in principle ideal for lattice calculations.\footnote{We notice that  leading piece of the  pNRQCD lagrangian  is renormalizable.}}

\subsubsection{Weakly coupled pNRQCD}

The lowest levels of quarkonium, like $J/\psi$, $\Upsilon (1S),\Upsilon (2S) \dots$,
may be described by weakly coupled pNRQCD, while the radii of the excited states 
are larger and presumably need to be described 
by strongly coupled pNRQCD. All this is valid for states away from strong-decay threshold, 
i.e. the threshold for a decay into two heavy-light hadrons. 
The effective Lagrangian is organized as an expansion in $1/m$  and $\als(m)$, 
inherited from NRQCD, and an expansion in $r$ (multipole expansion) \cite{Brambilla:1999xf}: 
\begin{equation}
 L_{\rm pNRQCD}  =    \int d^3R\,       \int d^3r\,  
\sum_n \sum_k \frac{c_n(\als(m),\mu)}{m^{n}}  
 V_{n,k}(r,\mu^\prime, \mu) \; r^{k}  
\times O_k(\mu^\prime,mv^2,...) ,
\end{equation}
where ${\bf R}$ is the center of mass position and 
 $O_k$ are the operators of pNRQCD. The matrix elements of these operators 
depend on the low-energy scale
$mv^2$ and $\mu'$, where $\mu^\prime$ is the pNRQCD factorization scale.  The $V_{n,k}$
are the Wilson coefficients of pNRQCD that encode the contributions 
from the scale $r$ and are nonanalytic in $r$. The $c_n$ are the NRQCD matching coefficients 
as given in (\ref{EQ:NRQCD}).

The degrees of freedom, which are relevant below the soft scale, and which 
appear in the operators $O_k$, are $Q\overline{Q}$ states (a color-singlet $S$ and a color-octet $O=O_a T^a$  state,
depending on  ${\bf r}$ and ${\bf R}$)
and (ultrasoft) gluon fields, which are expanded in $r$ as well (multipole expanded and depending only on ${\bf R}$).
pNRQCD makes apparent that the correct zero order problem is the Schr\"odinger equation.
Looking at the equations of motion of pNRQCD, we may identify 
$V_{n,0}= V_n$ with the $1/m^n$ potentials that enter 
the Schr\"odinger equation and 
$V_{n,k\neq 0}$ with the couplings of the 
ultrasoft degrees of freedom, which
provide corrections to the Schr\"odinger equation.
 Nonpotential interactions, 
associated with the propagation of low-energy degrees of freedom
are, in general, present as well, and start 
to contribute at NLO in the multipole expansion.
{\it They are typically related to nonperturbative effects and are carried 
by purely gluonic correlator local or nonlocal in time: they need to be calculated on  the lattice.}
pNRQCD enables precise and systematic higher order calculations on bound state allowing the extraction of 
precise determinations of standard model parameters like the quark masses and $\alpha_s$  from quarkonium.

\subsection{Strongly coupled pNRQCD}

\begin{figure}[!tbp]
 \begin{minipage}[b]{0.32\textwidth}
    \includegraphics[width=\textwidth]{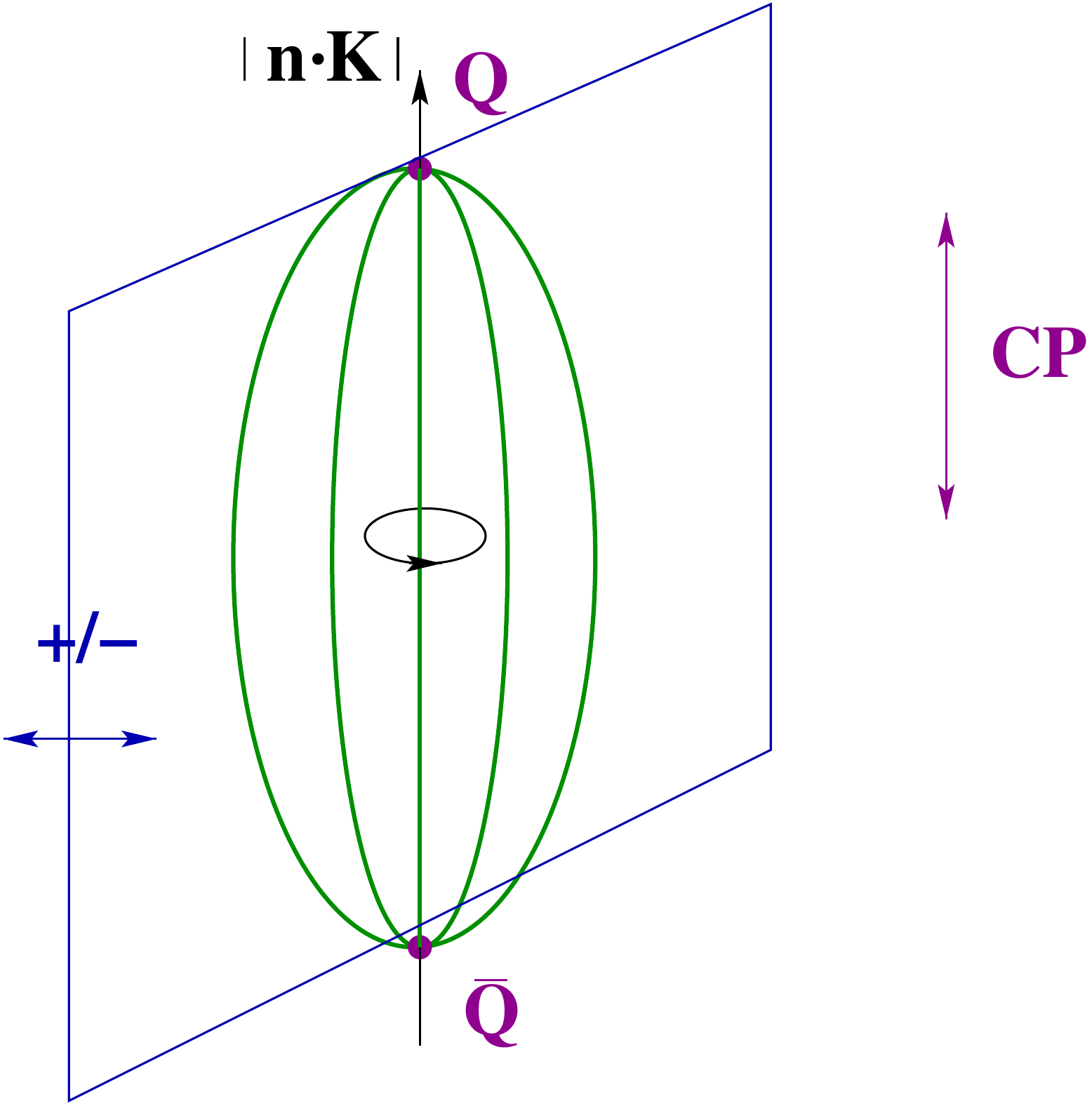}  
    \vskip 1.2 truecm
    \caption{Quarkonium hybrid symmetries.} \label{figsym}
     \end{minipage}
\hfill
  \begin{minipage}[b]{0.32\textwidth}
    \includegraphics[width=\textwidth]{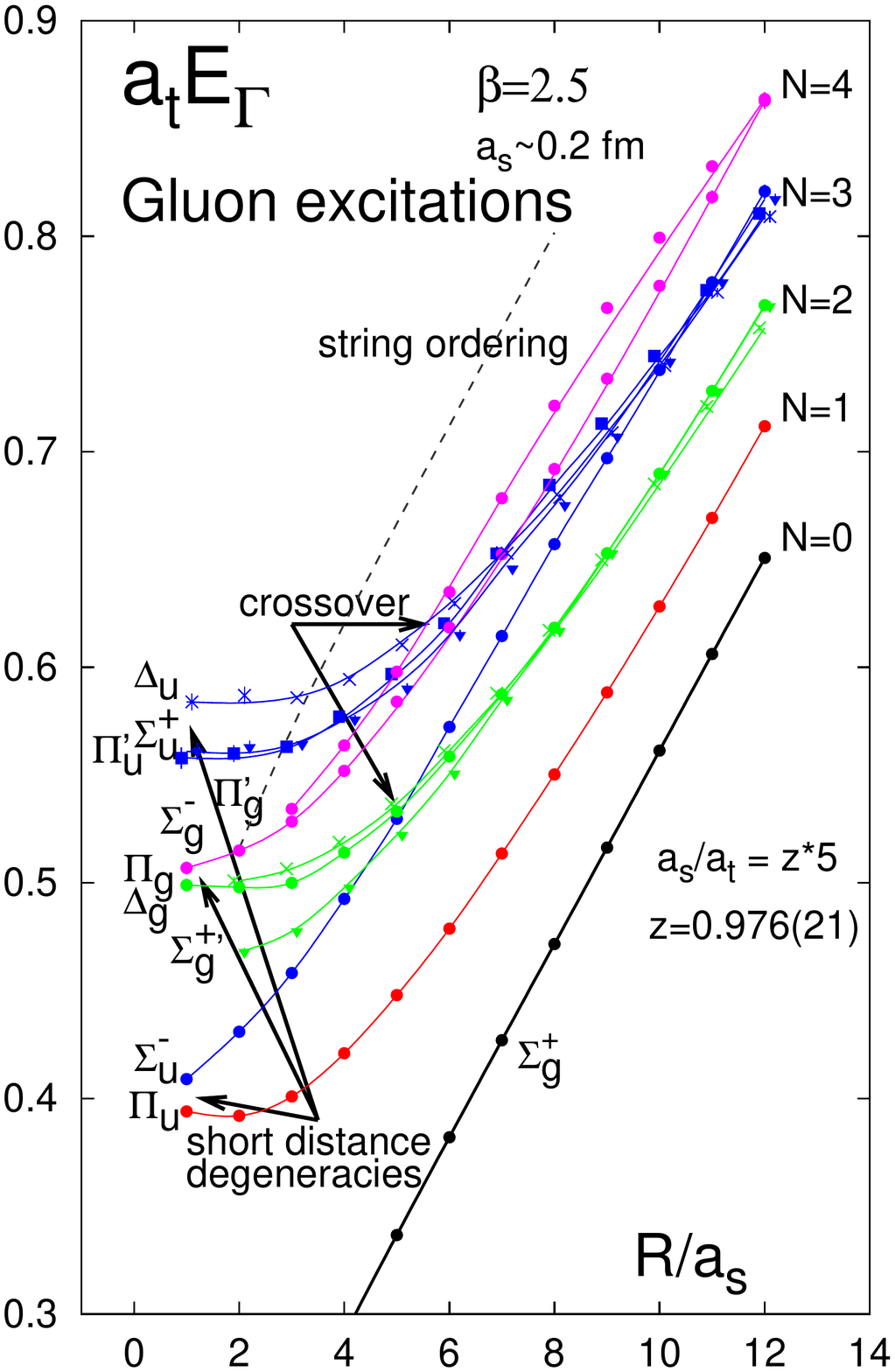} 
    \caption{Hybrid static energies, $E_\Gamma$, in lattice units, from~\cite{Juge:2002br}.}
      \label{figEg}
 \end{minipage}      
\end{figure}

When $mv \sim \lQ$  the soft scale is nonperturbative, the matching cannot be performed in 
perturbation theory any more. Only singlet degrees of freedom exist and they include $Q\bar{Q}$
states, hybrids $Q\bar{Q} g$ states and glueballs. {\it Since the physics is nonperturbative
we need to use {\bf lattice input} to construct the EFT}.
In particular we need the lattice evaluation of the gluonic static energies of  $Q\bar{Q}$ pair: the have been calculated on the lattice since
long \cite{Foster:1998wu,Juge:2002br,Bali:2003jq}
and recently updated in \cite{Muller:2019joq, Schlosser:2021wnr}
and they use generalized Wilson loops.  The gluonic static energies, $E_\Gamma$ in Fig.~\ref{figEg},
are classified according to representations of the symmetry group $D_{\infty\,h}$, typical of molecules, 
and labeled by $\Lambda_\eta^\sigma$ (see Fig.~\ref{figsym}):
$\Lambda$ is the rotational quantum number $|\hat{\bf r}\cdot{\bf K}| = 0,1,2,\dots$,
with ${\bf K}$ the angular momentum of the gluons, 
that corresponds to $\Lambda = \Sigma, \Pi, \Delta, \dots$;
$\eta$ is the CP eigenvalue ($+1\equiv g$ (gerade) and $-1 \equiv u$ (ungerade));
$\sigma$ is the eigenvalue of reflection with respect to a plane passing through the $Q\bar{Q}$ axis.
The quantum number $\sigma$ is relevant only for $\Sigma$ states.
In general there can be more than one state for each irreducible representation of $D_{\infty\,h}$: 
higher states are denoted by primes, e.g., $\Pi_u$, $\Pi_u'$, $\Pi_u'', \dots$
{\it This set of static energies will be fundamental to address the exotics.}

Since at this moment we are dealing with $Q\bar{Q}$ states below threshold we are interested merely in
$\Sigma_g^+$ static energy (that in this case coincides with the static singlet $Q\bar{Q}$ potential)
and in the information that such curve develops a gap of order $\lQ$ at a distance $r\sim \lQ^{-1}$: therefore
all the hybrid static energies can be integrated out as pNRQCD follows from integrating out all degrees of freedom
with energy up to $mv^2$.
The quarkonium singlet field $S = \bar{Q}Q$ 
 is the only low-energy dynamical 
degree of freedom in the pNRQCD Lagrangian (up to US pions)
which reads \cite{Brambilla:2000gk,Pineda:2000sz,Brambilla:2004jw}:
\begin{equation}
\quad   L_{\rm pNRQCD}=   \int d^3R\,       \int d^3r\, 
{ S}^\dagger
   \left(i\partial_0-\frac{{\bf p}^2}{2m}-V_S(r)\right){S}\, 
 \end{equation}
 and lends support to potentials models in this regime, with the difference that
 the singlet potential $V_s(r) = V_0 + V_1/m +V_2/m^2$  is the QCD potential
 calculated in the matching \cite{Brambilla:2000gk,Pineda:2000sz}
 and appear to have differences with respect to the phenomenological potential models.
 The potential at order $1/m^2$ contains a velocity dependent and a spin dependent part, with a spin orbit,
 spin-spin and spin tensor interaction. {\it Notice that spin effects are appearing only at order $1/m^2$ and are
 therefore suppressed in the spectrum and in the transitions: we will see that things are different for exotics.}
{\it These potentials  are  nonperturbative quantities in the form of expectation values of generalized (with
chromolectric and chromagnetic insertions) gauge invariant 
Wilson-loop operators to be evaluated on the lattice. Actually it is the EFT that lends a clean definition and an interpretation
of the static Wilson loops measured on the lattice as actual potentials in this regime, together with a prescription to use them
to calculate observables.
Some of these generalized Wilson loops  have been calculated on the  lattice
(only quenched) \cite{Bali:1997am,Bali:2000gf,Koma:2006si,Koma:2012bc,Koma:2010zza}
but some contributions at order $1/m^2$ are still to be calculated.}
Using these potentials,  all the masses for heavy quarkonia away from threshold
can be obtained by the solution of the Schr\"odinger equation
with such potentials. Lorentz invariance is still there in the form if exact relations among potentials and it has been observed on the lattice.
Decays are described by calculating the imaginary parts of the potentials \cite{Brambilla:2002nu} where nonperturbative contributions
enter in the form of gauge invariant time non local chromoelectric and chromomagnetic correlators that have still to be calculated on the lattice.
Summarizing, strongly coupled pNRQCD
factorizes low energy nonperturbative contributions in terms of generalised gauge invariant Wilson loops opening the way
to a systematic study of the confinement mechanism and systematic applications to quarkonium spectrum and decay.

\subsubsection{Quarkonium production}

NRQCD is the theory used to study quarkonium production, however one of the main problem  is the proliferation
of LDMEs, i.e. the nonperturbative matrix elements of singlet and octet four fermions operators:
they are flavor dependent and increase in number when one seeks precision in the expansion in $v$.
While those containing singlet operators can be reduced to the wave functions, the octet ones had to be extracted
from the data, with contradictory results. The lower  energy factorization obtained in pNRQCD may have  a great
impact also on quarkonium production: it has been shown 
that,  for the case of inclusive quarkonium production in P wave,  the octet LDMEs can be factorized in the product
of the quarkonium wave function derivative in the origin and a  chromoelectric correlator \cite{Brambilla:2020ojz,Brambilla:2021abf}
\begin{equation} 
\label{eq:hqoctet} 
\langle \Omega  |{\cal O}^{h_Q} ({}^1S_0^{[8]})  \Omega \rangle = 3 \times \frac{3 N_c}{2 \pi} |R^{(0)}{}'(0)|^2 \frac{1}{9 N_c m^2} {\cal E}, 
\end{equation} 
where ${\cal E}$ is the dimensionless gluonic correlator 
\begin{equation} 
\label{eq:ecorrelator} 
{\cal E} = \frac{3}{N_c} \int_0^\infty dt\, t \; \int_0^\infty dt'\, t' \;
\langle \vert \Omega | \Phi_\ell^{\dag ab} \Phi_0^{\dag ad} (0;t) g {E}^{d,i}(t) g {E}^{e,i}(t') \Phi_0^{ec} (0;t') \Phi_\ell^{bc} | \Omega \rangle.  
\end{equation} 
where  $\Phi_\ell$ is a Wilson line along the (arbitrary) direction  $\ell$  (up to infinity) in the adjoint representation and 
$ \vert \Omega \rangle$ is the QCD vacuum state.
This is similar to what one gets in the decay factorization of the NRQCD matrix elements \cite{Brambilla:2002nu}
\begin{equation} 
\label{eq:hqoctet_decay} 
\langle h_Q | 
\psi^\dag T^a \chi \chi^\dag T^a \psi 
| h_Q \rangle 
= \frac{3 N_c}{2 \pi} |R^{(0)}{}'(0)|^2 \frac{1}{9 N_c m^2} {\cal E}_3, 
\end{equation} 
where the correlator ${\cal E}_3$ is defined by 
\begin{equation} 
\label{eq:E3def} 
{\cal E}_3 = \frac{1}{2 N_c} 
\int_0^\infty t^3\, dt \langle \Omega | 
g {E}^{a,i}(t) 
\Phi_0^{ab} (t,0) 
g {E}^{b,i}(0) | \Omega \rangle.  
\end{equation} 
{\it Hence the pNRQCD expressions have more predictive power than the corresponding NRQCD
ones. In particular, correlators determined with charmonium data may be used to compute color
octet matrix elements and hence observables in the bottomonium sector. }

\subsection{Quarkonium in medium}

pNRQCD has been extended to the case of finite temperature $T$ 
which allows to calculate  in the weak coupling the finite $T$  quarkonium interaction potential and the thermal effects to the energies
and widths \cite{Brambilla:2020esl}.  In the strong coupling the potential is obtained from a lattice calculation of the Wilson loop at finite $T$
\cite{Bala:2021fkm}.
In particular combining pNRQCD and open quantum system \cite{Brambilla:2017zei},  it has been possible to describe the 
nonequilibrium evolution of small quarkonia systems (bottomonium)  inside the strongly coupled Quark Gluon Plasma
in the hierarchy $m \gg 1/r\gg T, m_D \gg mv^2$  with an evolution equation for the singlet and octet 
density matrix of the Linblad type  \cite{Brambilla:2020qwo}.
The interesting thing is that the properties of the strongly coupled QGP are carried 
just by two transport coefficients that are appropriate correlators of electric fields at finite $T$ \cite{Brambilla:2017zei}:

\begin{equation}  
	\kappa=\frac{g^{2}}{6N_{c}} {\rm Re} \int^{\infty}_{-\infty} \,\mathrm{d}s~\Big \langle {\rm T} 
	E^{a,i}(s)   \phi^{ab}(s,0)   E^{a,i}(0)  \Big \rangle, 
	\gamma=-\frac{g^{2}}{6N_{c}}   {\rm Im}     
	\int^{\infty}_{\infty} \, \mathrm{d}s~\Big \langle {\rm T} 
	E^{a,i}(s)   \phi^{ab}(s,0)   E^{a,i}(0)  \Big \rangle, 
	  \label{eq:gamma_def}
        \end{equation}
        where $\langle \dots \rangle$ is a shortcut to indicate the QCD average at finite T on the QCD vacuum
        $\vert \Omega\rangle$ (and  ${\rm T}$ is time ordering.) 
The quantity $\kappa$ is the heavy quark momentum diffusion coefficient.
$\kappa$ has been  evaluated on the lattice \cite{Altenkort:2020fgs}
for a medium   locally in equilibrium at a temperature $T $.
In particular, it is important to obtain their $T$ dependence in a large range  of $T$ as done in \cite{Brambilla:2020siz}.
Using the EFT one can relate these transport coefficients to the thermal modification of the energy levels and to the
thermal widths of quarkonium,
which  allows us to use unquenched lattice calculations
\cite{Aarts:2011sm,Kim:2018yhk}  to determine them \cite{Brambilla:2019tpt}.
 
This formalism can be extended to describe the evolution of exotic states inside the QGP created in heavy ion collisions.

\subsection{Role of the Lattice and Input needed from Lattice}

Summarizing, for what concerns quarkonium below threshold, lattice results have been   key in order 
to  formulate strongky coupled pNRQCD and lattice input  is still needed, precisely,
we need:
\begin{itemize}
\item{}Lattice calculation of  chromoelectric and chromomagnetic  correlators local in time and nonlocal in time: they contribute to 
quarkonia spectra and decay \cite{Brambilla:2004jw,Brambilla:2002nu}.
\item{}Lattice calculation of generalized static Wilson loops with insertions of chromoelectric and chromomagnetic fields: they contain the information 
on the nonperturbative singlet static potential and its relativistic corrections.
\item{}Lattice calculation of the chromolelectric correlators appearing in production: they allow to calculate the LDMEs  necessary to describe 
quarkonium production. They are somehow similar to what is done for TMDs and PDFs.
\item{}Lattice calculation of the transports coefficients: finite $T$ lattice calculations of such objects in  a wide range of $T$ will enable us 
to use the lattice to study nonequilibrium processes   \cite{Brambilla:2017zei,Brambilla:2020qwo,Brambilla:2020siz}. There is not a yet 
a direct evaluation of $\gamma$.
\end{itemize}
{\it This will be instrumental to sharpen  our control of quarkonium spectrum, decays, production and propagation in medium and {\bf to prepare 
for the treatment of the exotics.}}

\section{Exotics at or above the strong decay threshold}
\begin{figure}[!tbp]
 \begin{minipage}[b]{0.4\textwidth}
    \includegraphics[width=\textwidth]{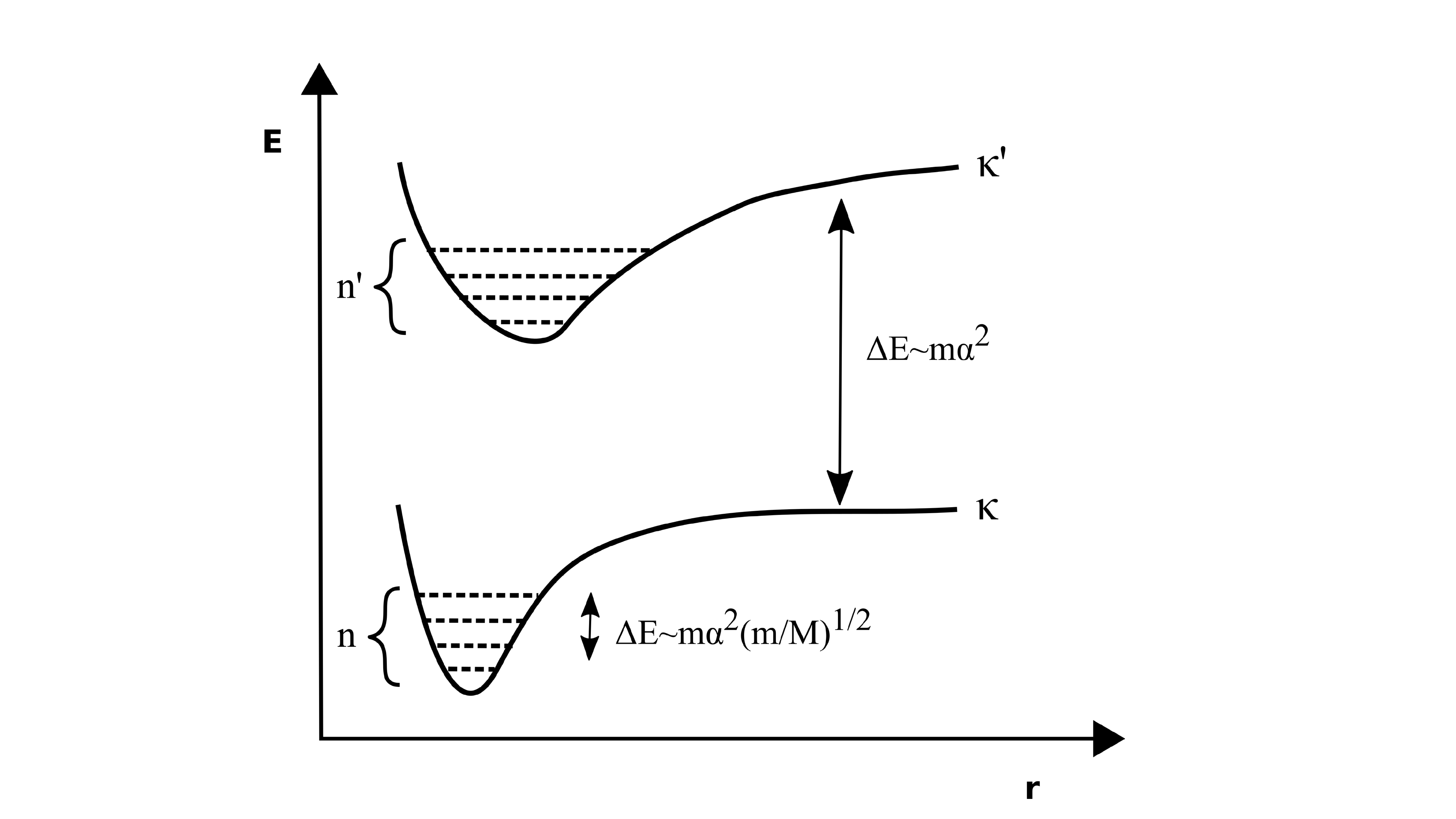}
    \vskip 0.2truecm  
    \caption{Electronic static energies in QED} \label{bo1}
     \end{minipage}
\hfill
  \begin{minipage}[b]{0.4\textwidth}
    \includegraphics[width=\textwidth]{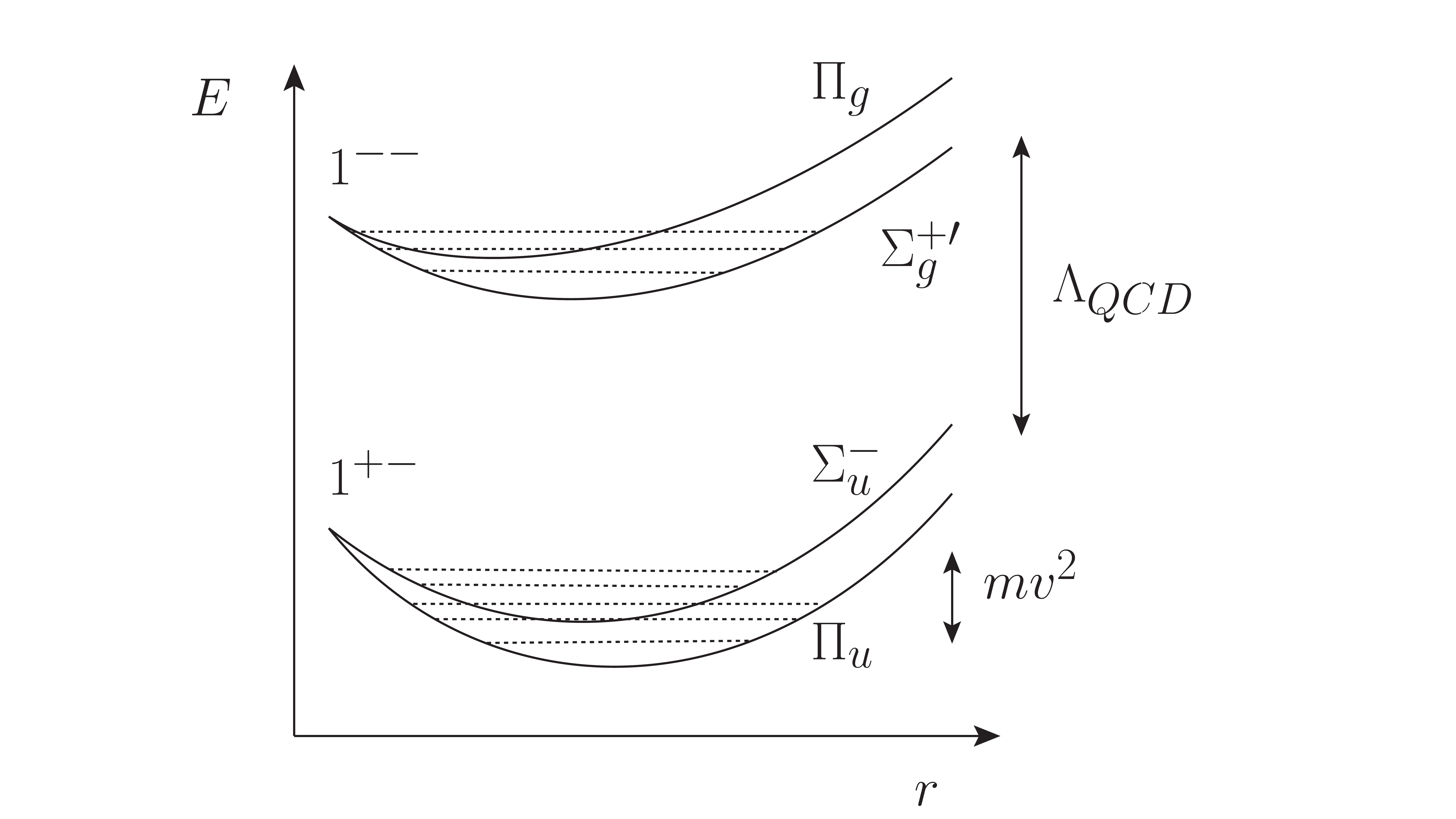} 
    \caption{Gluonic (or Hybrid) static energies, $E_\Gamma$, in QCD.}
      \label{bo2}
 \end{minipage}      
\end{figure}

In the most interesting region,  close or above the strong decay  threshold, where the X Y Z have been discovered,
 the situation is much more complicated: 
 there is no mass gap between quarkonium and the creation of a heavy-light mesons couple, nor to gluon 
 excitations, therefore  many additional states built on the light quark quantum numbers may appear.
 The threshold region remains troublesome
also for the lattice,  although  several ab initio calculations of the exotic masses have been recently being pionereed.
Still, $m$  is a  large scale and NRQCD is applicable. Then, if we want to introduce a description 
 of the bound state similar to pNRQCD, making apparent that the zero  order problem is the Schr\"odinger equation,
 we can still count on  scale separation.
  Let us consider   bound states of two nonrelativistic particles and some light d.o.f., e.g. molecules in QED  or quarkonium 
  hybrids ($Q\bar{Q} g$ states) or tetraquarks  ($Q\bar{Q}   q\bar{q}$  states) in QCD:
electron/gluon fields/light quarks  change adiabatically in the presence of heavy quarks/nuclei.  The heavy quarks/nuclei interaction may be described at leading 
order in the nonrelativistic expansion by an effective  potential $V_\kappa$ between the static sources where $\kappa$ labels different excitations 
of the light degrees of freedom.  A plethora of states  can be  built on each on the potentials $V_\kappa$  by solving the
corresponding Schr\"odinger equation, see Figs.~\ref{bo1} and \ref{bo2}.  
This picture corresponds to the Born-Oppenheimer (BO) approximation. Starting from pNRQED/pNRQCD the BO approximation can be made 
rigorous and cast into a suitable EFT called Born-Oppenheimer EFT (BOEFT)  \cite{Berwein:2015vca,Brambilla:2017uyf,Oncala:2017hop,Soto:2020xpm,Brambilla:2019jfi,Brambilla:2018pyn}
which exploits the hierarchy of scales
$\lQ \gg mv^2$ (the analogous  in QED of  the energy of the electrons being greater than the energy of nuclei or
conversely the typical time 
of the electrons (fast degrees of freedom) being bigger than the typical time of the  nuclei (slow degrees of freedom)).
{\it BOEFT  allows to address all exotics pictures (hybrids, tetraquarks, molecules and pentaquarks) in the same framework.
Of course in QCD the scale $\lQ$ is nonperturbative and we need to use the lattice to calculate the appropriate gluonic static energies, see Fig.  \ref{figEg}.}

\subsection{Hybrids: Born Oppenheimer Effective Field Theory}

In the following we consider the example of hybrids, i.e. states made by $Q\bar{Q}g$.
We  consider hybrids static energies that are
labelled by the quantum numbers $\Lambda^\sigma_\eta$ , see Sec. 2.4 and Fig. 2 and
we restrict to considering the two  lowest-lying static energies ones $\Pi_u$ and $\Sigma_u^-$.
Modes associated to higher potentials are separated from the $\Pi_u$ and $\Sigma_u^-$ potentials by an energy gap of order $\lQ$
and are integrated out with that scale.
In the $r \to 0$ limit $\Pi_u$ and $\Sigma_u^-$ are degenerate and correspond to a gluonic operator with quantum numbers~$\kappa \equiv 
K^{PC}=1^{+-}$,  ${\bf K}$ being the gluon angular momentum,
because the group $D_{\infty\,h}$ becomes the more symmetric group O(3)$\times$C \cite{Brambilla:1999xf,Brambilla:2004jw,Foster:1998wu}.
We recall that  $\lambda$ is the quantum number $ \lambda \equiv {\bf K} \cdot \hat{r}; \vert \lambda \vert =\Lambda$.

The hybrid static energies are defined by ($\Gamma$ stays for $\Lambda^\sigma_\eta$)
\begin{equation}
  E_\Gamma(r)  =  \lim_{T\to \infty} {i \over T}  \log \langle X_\Gamma, T/2\vert X_\Gamma , -T/2\rangle  ; \quad 
  \vert X_\Gamma \rangle =  \chi ({\bf x}_2)\phi({\bf x}_2,
                   {\bf R}) T^a P_\Gamma^a( {\bf R}) \phi({\bf R}, {\bf x}_1) \psi^\dagger({\bf x}_1)\vert \Omega\rangle
                   \label{gamma}
                 \end{equation}
                 where $P^a_\Gamma$ is some gluonic operator that generates the right quantum numbers $\Gamma$.
                 For our case we have $P^a_\Gamma = \hat{\bf r} \cdot {\bf B}^a, \hat{\bf r}  \times {\bf B}^a$ respectively for the
                 $\Sigma_u^-$ and the $\Pi_u$, ${\bf B}^a$ being the chromomagnetic field.
In the short-interquark-distance limit $r\rightarrow 0$, quarkonium hybrids reduce to
gluelumps, which are color-singlet combinations of a local static octet color $Q\bar{Q}$ source coupled to a gluonic field \cite{Brambilla:1999xf,Berwein:2015vca}.
We define the gluelump operators, ${G}^{ia}_{\kappa}$, as the Hermitian color-octet operators that generate the eigenstates of
the zero order NRQCD Hamiltonian  $H_{0}$ in the presence of a local heavy-quark-antiquark octet source:
\begin{align}                                                                                                                
H_0 {G}^{ia}_{\kappa}(\bm{R},t)|0\rangle &= \Lambda_{\kappa} {G}^{ia}_{\kappa}(\bm{R},t)|0\rangle\,,                             
\end{align}
where $a$ is the color index and $i$ the spin component (it labels the
component of an  irreducible representation of $SO(3)$ with Casimir K),
 $\kappa$ labels the quantum numbers $K^{PC}$ of the gluonic degrees of freedom. For our case $ {\bf G}^{a}=  {\bf B}^{a}$
with $K^{PC}=1^{+-}$.
The spectrum of the mass eigenvalues, $\Lambda_{\kappa}$, called the gluelump mass, has been computed on the lattice
in Refs.~\cite{Foster:1998wu,Bali:2003jq,Marsh:2013xsa} and it is given by the gluelump correlator
$
\Lambda_{\kappa} =\lim_{T\to \infty} {i/T} \ln \langle \Omega \vert G^{a}_{\kappa}(T/2) \phi_{ab}(T/2, -T/2) 
G^{b}_{\kappa}(-T/2)\vert \Omega \rangle$.

The static energy $E_\Gamma$ calculated in NRQCD at small r  can be matched to the following objects
in BOEFT:
\begin{equation}
E_\Gamma \equiv E_{\kappa, \lambda} = V^{(0)}_{\kappa,\lambda}= {\alpha_s\over 6 r} + \Lambda_\kappa +a_{\kappa \lambda} r^2 +\cdots
\end{equation}
where $\alpha_s/6 r$ is the static octet potential,
$a_{\kappa \lambda}$ is a nonperturbative coefficient that has a field theoretical definition and starts
to depend on $\lambda$ (instead that only on $\kappa$), which makes clear that 
at the NLO order in the multipole expansion the system is no longer spherically symmetric
but acquires instead a  cylindrical symmetry around the heavy-quark-antiquark axis.
Therefore it is convenient to work with a basis of states with good transformation properties under $D_{\infty  h}$. Such states 
can be constructed by projecting the gluelump operators on various directions with respect to the heavy-quark-antiquark axis.
We define the degrees of freedom of the BOEFT as the operator
$\hat{\Psi}_{\kappa\lambda}(\bm{r},\,\bm{R},\,t)$ defined by
\begin{equation}                                                                                                                                             
  P^{i\dag}_{\kappa\lambda} O^{a}\left(\bm{r},\bm{R},t\right) G_{\kappa}^{ia}(\bm{R},t)
  =Z_\kappa^{1/2} \hat{\Psi}_{\kappa\la}(\bm{r},\bm{R},t)\,,\label{eq:relation}                                                    
\end{equation} the 
  $Z$ is a field renormalization,
$P^i_{\kappa\lambda}$ is a projector that projects the gluelump operator to an eigenstate
of $\bm{K}\cdot\hat{\bm{r}}$ with eigenvalue $\lambda$.
For our case the projectors $P^{i}_{\lambda}$ read 
$
P^{i}_{0}=\hat{r}_0^i= \hat{r}^i\,,
P^{i}_{\pm 1}=\hat{r}^i_{\pm}=\mp\left(\hat{\theta}^i\pm i\hat{\phi}^i\right)/\sqrt{2}\,, \label{p}
$
with
$
\hat{\bm{r}}=(\sin(\theta)\cos(\phi),\,\sin(\theta)\sin(\phi)\,,\cos(\theta))$,
$\hat{\bm\theta}=(\cos(\theta)\cos(\phi),\,\cos(\theta)\sin(\phi)\,,-\sin(\theta))$,
$\hat{\bm\phi}=(-\sin(\phi),\,\cos(\phi)\,,0)$.
The BOEFT is obtained by integrating out modes of scale $\Lambda_{\rm QCD}$, i.e. the gluonic excitation
and the Lagrangian reads as
\begin{equation}
  L_{\rm BOEFT} = \int d^3Rd^3r \, \sum_{\kappa} \sum_{\lambda\lambda^{\prime}}
  \textrm{Tr}\{ \hat{\Psi}^{\dagger}_{\kappa\lambda}(\bm{r},\,\bm{R},\,t) [i\partial_t - V_{\kappa\lambda\lambda^{\prime}}(r)+
  P^{i\dag}_{\kappa\lambda}\frac{\bnabla^2_r}{m}P^{i}_{\kappa\lambda^{\prime}}]\hat{\Psi}_{\kappa\lambda^{\prime}}(\bm{r},\,\bm{R},\,t)\}
                +\dots \,,                                             
\label{bolag2}                                                                                                                                   
\end{equation}
where the trace is over spin indices of the heavy quark and antiquark, and the ellipsis stands for operators producing transitions
to standard quarkonium states and transitions between hybrid states of different $\kappa$. 
The latter are suppressed at least by $1/\Lambda_{\rm QCD}$ since the static energies for different $\kappa$
are separated by a gap $\sim\Lambda_{\rm QCD}$. The interesting points of the Lagrangian above is that the kinetic operator is not commuting
with the projector, which induces a non adiabatic coupling, already known from molecular physics.

The potential $V_{\kappa\lambda\lambda^{\prime}}$ can be organized into an expansion in $1/m$ and a sum
 of static,  spin-dependent (SD) and independent (SI) parts to be calculated in the matching with NRQCD.
 In Ref.~\cite{Berwein:2015vca} the static potential $V^{(0)}_{\kappa\lambda}(r)$ was matched to the quark-antiquark hybrid static energies computed on the lattice
 and the matrix elements of $P^{i\dag}_{\kappa\lambda}\frac{\bnabla^2_r}{m}P^i_{\kappa\lambda^{\prime}}$ were obtained for $\kappa=1^{+-}$
and  shown to contain off-diagonal terms in $\lambda$-$\lambda^{\prime}$ that lead to coupled Schr\"odinger equations:
\bea
  &&\hspace{-1cm}
  \left[-\frac{1}{Mr^2}\,\partial_rr^2\partial_r+\frac{1}{Mr^2}
  \begin{pmatrix} l(l+1)+2 & 2\sqrt{l(l+1)} \\
    2\sqrt{l(l+1)} & l(l+1)
  \end{pmatrix}
  +
  \begin{pmatrix} E_{\Sigma_u^-}    & 0 \\
    0 & E_{\Pi_u}
  \end{pmatrix}\right]\hspace{-4pt}
\begin{pmatrix} \psi_\Sigma^{(N)} \\
  \psi_{-\Pi}^{(N)}
\end{pmatrix}  
= \mathcal{E}_N
\begin{pmatrix} \psi_\Sigma^{(N)} \\
  \psi_{-\Pi}^{(N)}
\end{pmatrix}\,,
\label{eq1}\\
&&\hspace{-1cm}
\left[-\frac{1}{Mr^2}\,\partial_r\,r^2\,\partial_r+\frac{l(l+1)}{Mr^2}+ E_{\Pi_u} \right]\psi_{+\Pi}^{(N)}=\mathcal{E}_N\,\psi_{+\Pi}^{(N)}\,,
\label{eq2}
\eea
where $l$ is the total  orbital angular momentum quantum number sum ${\bf L} = {\bf L}_{Q\bar{Q}}+ {\bf  K }$, $\mathcal{E}_N $ are the hybrids
energies and  $E_{\Sigma_u^-}$ and  $E_{\Pi_u}$ have been written in BOEFT in term of the static potentials plus the gluelump mass taken from 
\cite{Bali:2003jq}: $\Lambda_{1^{+-}} =0.87(15)$ GeV\footnote{Note that the gluelump mass is a scheme dependent quantity like the quark mass, here it is
  calculated in RS scheme.}.

The
Schr\"odinger equations were solved numerically and the spectrum and wave functions of hybrid states generated the static energies labeled by $\Sigma^-_u$
and $\Pi_u$ were obtained, precisely:

The functions  $\psi_\Sigma^{(N)}$ and $\psi_{\pm \Pi}^{(N)}$ are radial wave functions;
$\psi_\Sigma^{(N)}$ and $\psi_{-\Pi}^{(N)}$ have negative parity and $\psi_{+\Pi}^{(N)}$ positive one.
The off-diagonal terms change the $\Sigma$ wave function to $\Pi$ and vice versa, but they do not change the parity.
Hence $\psi_\Sigma^{(N)}$ mixes only with $\psi_{-\Pi}^{(N)}$, and $\psi_{+\Pi}^{(N)}$ decouples.
For $l=0$ the off-diagonal terms vanish, so the equations for $\psi_\Sigma^{(N)}$ and $\psi_{-\Pi}^{(N)}$ decouple;
there exists only one parity state, and its radial wave function is given by a Schr\"odinger equation with the $E_\Sigma$ potential and an angular part $2/Mr^2$.
The eigenstates of the equations \eqref{eq1} and \eqref{eq2} are organized in the multiplets shown in Table~\ref{tabmultiplets} (all the $J^{PC}$ quantum numbers of a given multiplet will be degenerate up to when we will consider spin  dependent interactions in Sec. 3.1.2).

\begin{table}[ht]
\centerline{
 \begin{tabular}{|c|c|c|c|c|}
 \hline
        & $\,l\,$ & $J^{PC}\{s=0,s=1\}$       & $E_\Gamma$           \\  \hline
  $H_1$ & $1$     & $\{1^{--},(0,1,2)^{-+}\}$ & $E_{\Sigma_u^-}$, $E_{\Pi_u}$ \\
  $H_2$ & $1$     & $\{1^{++},(0,1,2)^{+-}\}$ & $E_{\Pi_u}$               \\
  $H_3$ & $0$     & $\{0^{++},1^{+-}\}$       & $E_{\Sigma_u^-}$          \\
  $H_4$ & $2$     & $\{2^{++},(1,2,3)^{+-}\}$ & $E_{\Sigma_u^-}$, $E_{\Pi_u}$ \\
  $H_5$ & $2$     & $\{2^{--},(1,2,3)^{-+}\}$ & $E_{\Pi_u}$               \\
  \hline
 \end{tabular}}
\caption{$J^{PC}$ multiplets with $l\leq2$ for the $\Sigma_u^-$ and $\Pi_u$ gluonic states. 
  The last column shows the gluonic static energies that appear in the Schr\"odinger equation of the respective multiplet.
\label{tabmultiplets}}
\end{table}

\subsubsection{Spectrum}

\begin{figure}[ht]
 \centering
  \begin{minipage}[b]{0.8\textwidth}
    \includegraphics[width=\textwidth]{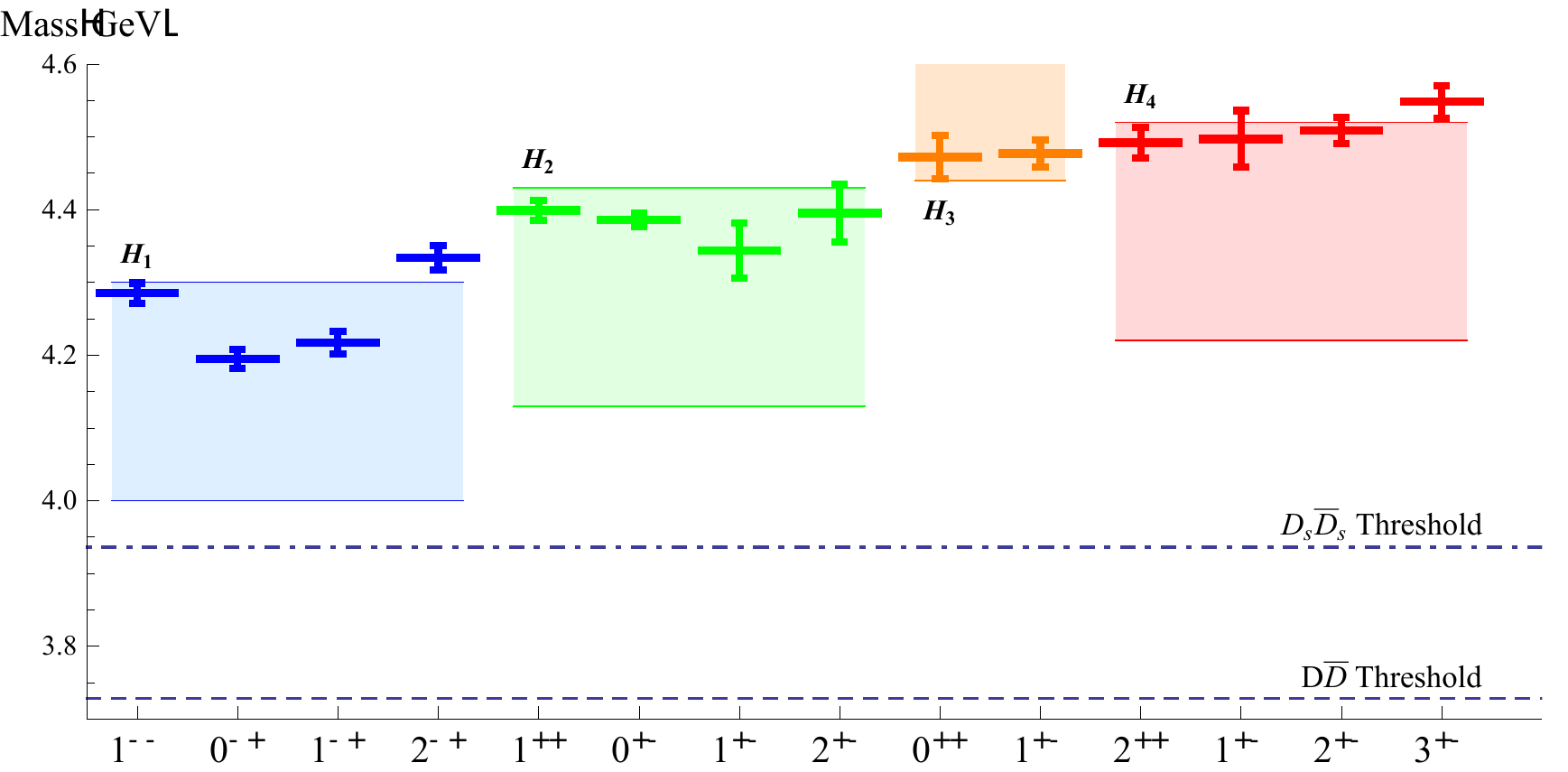}  
    \caption{Comparison of results from the direct lattice computations of charmonium hybrid masses of~\cite{HadronSpectrum:2012gic}
      with the results of~\cite{Berwein:2015vca}.
  The direct lattice mass predictions are plotted in solid lines with error bars corresponding to the mass uncertainties. Our results for the H1, H2, H3, and H4 multiplets have been plotted in error bands corresponding to the gluelump mass uncertainty of $\pm 0.15 $ GeV.
\label{figlatc}}
       \end{minipage}
\end{figure}

Keeping in the equations \eqref{eq1} and \eqref{eq2} only the heavy quark-antiquark kinetic energy and the hybrid static energies, $E_\Gamma$,
amounts at the {\em Born--Oppenheimer approximation}.
Keeping only the diagonal terms amounts at the {\em adiabatic approximation}.
The exact leading order equations include both diagonal and off-diagonal terms that define the so-called {\em non-adiabatic coupling}.
As it is clear from \eqref{eq1} and \eqref{eq2} both diagonal and off-diagonal terms contribute at the same order to the energy levels in the quarkonium hybrid case.
This situation is different from the case of (electromagnetic) molecules, where the non-adiabatic coupling is subleading~\cite{Brambilla:2017uyf}.
A physical consequence of the mixing is the so-called {\it $\Lambda$ doubling}, i.e., the lifting of degeneracy between
states with the same parity. We show this effect in Fig.~\ref{figlatc}, where you see that the multiplets $H_1$ and $H_2$ are not degenerate and this effect is also confirmed in direct lattice evaluations.
The same effect would be  seen also between $H_4$ and $H_5$ \cite{Berwein:2015vca}.
The effect is also present in molecular physics, however, $\Lambda$ doubling is a subleading effect there.

\begin{figure}[ht]
 \centering
  \begin{minipage}[b]{0.8\textwidth}
    \includegraphics[width=\textwidth]{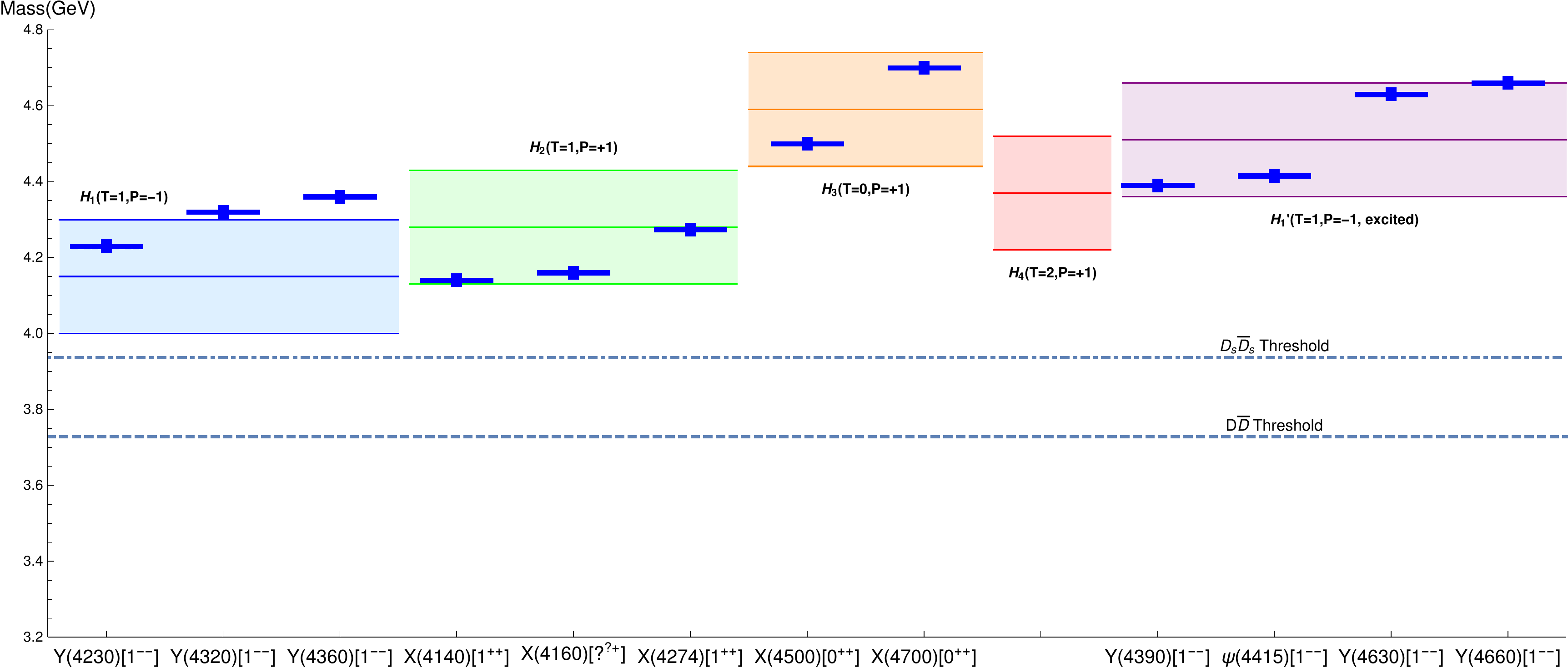}  
    \caption{Mass spectrum of neutral exotic charmonium states obtained by solving Eqs.    \eqref{eq1} and \eqref{eq2}.
     The experimental states of Sec. 3.1 that have matching quantum numbers are plotted in solid blue lines. In the figure $T$ stay for the total angular momentum. $H_1^\prime$ is the first $H_1$ radial excitation of $H_1$.
     The multiplets have been plotted with error bands corresponding to a gluelump mass uncertainty of 0.15 GeV. Figure taken from \cite{Brambilla:2019esw}.} \label{figexp}
     \end{minipage}
\end{figure}

Several charmonium-like states have been found by the B factories in the last decade (for a recent review see~\cite{Brambilla:2019esw}).
We compare some of these states with the hybrid spectrum obtained from solving the coupled Schr\"odinger equations \eqref{eq1} and \eqref{eq2} in Fig.~\ref{figexp}.
In the bottomonium sector, the state $Y_b(10890)[1^{--}]$ with mass $M_{Y_b} = (10.8884 \pm 3.0)$~GeV found by BELLE may be a possible $H_1$ candidate,
for which we find $M_{H_1} = (10.79 \pm 0.15)$~GeV.
{\it This picture allows us to describe the spectrum of hybrids within an EFT description of QCD (which should be equivalent to QCD in 
the taken window) plus lattice input. It is complementary to direct lattice evaluations  of the spectrum in many ways:
first of all it is simpler and allows to get all the flavour of hybrids on the basis of the evaluation of few correlators only dependent on the gluonic 
degrees of freedom, it takes advantage of factorization and it lends an interpretation to the underlying physics.
The multiplets that we have obtained are distinctive of the underlying physics: if one would construct the hybrids multiplets using the constituent  gluon picture  for example a
different multiplet structure would be obtained  \cite{Berwein:2015vca}.
Of course in order to validate the experimental  identifications one has to add the spin structure, and especially consider decays and production features:
also this may turn out to  be simpler  in the BOEFT approach.}

\subsubsection{Spin effects in Hybrids}

The spin of the quarks has been considered in  direct lattice calculations of the hybrids mass, mostly
in charmonium  for quenched and  noncontinuum extrapolated data \cite{Cheung:2016bym}, recently in  unquenched and continuum extrapolated  data \cite{Ray:2021nhe}
 and recently in bottomonium \cite{Ryan:2020iog}.
 It is rare to see spin effects treated in models and, in case they are considered,  it is on the basis of a gluon constitutent
 picture or extrapolating from the 
standard quarkonium spin structure, which we will see is not the appropriate thing to do.
In BOEFT we could obtain for the first time the spin dependent potentials (for  $\kappa=1^{+-}$,
$\lambda$ takes the values $0,\pm 1$)  \cite{Brambilla:2019jfi,Brambilla:2018pyn,Soto:2020xpm}:
\begin{align}
V_{\la\lap\,SD}^{(1)}(r)&=V_{SK}(r)\left(P^{i\dag}_{\la}\bm{K}^{ij}P^j_{\lap}\right)\cdot\bm{S}\nonumber\\
&\quad + V_{{SK}b}(r)\left[\left(\bm{r}\cdot \bm{P}^{\dag}_{\la}\right)\left(r^i\bm{K}^{ij}P^j_{\lap}\right)\cdot\bm{S}
-\left(r^i\bm{K}^{ij}P^{j\dag }_{\la}\right)\cdot\bm{S} \left(\bm{r}\cdot \bm{P}_{\lap}\right)\right]
\,,\label{sdm}\\
V_{\la\lap\,SD}^{(2)}(r)&=V_{SLa}(r)\left(P^{i\dag}_{\la}\bm{L}_{Q\bar{Q}}P^i_{\lap}\right)\cdot\bm{S}+V_{SLb}(r)P^{i\dag}_{\la}\left(L_{Q\bar{Q}}^iS^j+S^iL_{Q\bar{Q}}^j\right)P^{j}_{\lap}\nonumber\\
&\quad + V_{SLc}(r)\left[\left(\bm{r}\cdot \bm{P}^{\dag}_{\la}\right)\left(\bm{p}\times\bm{S}\right)\cdot \bm{P}_{\lap}
+\bm{P}^{\dag }_{\la}\cdot\left(\bm{p}\times\bm{S}\right)\left(\bm{r}\cdot \bm{P}_{\lap}\right)\right] \nonumber\\
&\quad +V_{S^2}(r)\bm{S}^2\de_{\la\lap}+V_{S_{12}a}(r)S_{12}\de_{\la\lap}+V_{S_{12}b}(r)P^{i\dag}_{\la}P^j_{\lap}\left(S^i_1S^j_2+S^i_2S^j_1\right)
\,,\label{sdm2}
\end{align}
where $\left(K^{ij}\right)^k=i\epsilon^{ikj}$ is the angular momentum operator for the spin-$1$ gluonic excitation
and $\bm{L}_{Q\bar{Q}}$ is the orbital angular momentum of the heavy-quark-antiquark pair,
${\bf S}$ is the total spin of the quarks, and the projectors for $\kappa=1^{+-}$ have been previously defined.
The $1/m$ operators in Eq.~(\ref{sdm}), with coefficients $V_{SK}(r)$ and $V_{SKb}(r)$
couple the angular momentum of the gluonic excitation with the total spin of the heavy-quark-antiquark pair.
These operators are characteristic of the hybrid states and are absent for standard quarkonia. 
Among the $1/m^2$ operators in Eq.~(\ref{sdm2}), the operators with coefficients $V_{SLa}(r)$, $V_{S^2}(r)$, and $V_{S_{12}a}$ are the standard spin-orbit, total spin squared, and tensor spin operators respectively, which appear for standard quarkonia also. In addition to them,
three novel operators appear at order $1/m^2$. The operators with coefficients $V_{SLb}(r)$ and $V_{SLc}(r)$
are generalizations of the spin-orbit operator to the hybrid states. Similarly,
the operator with coefficient $V_{S_{12}b}(r)$ is generalization of the tensor spin operator to the hybrid states. 
{\it Quite interestingly, differently from the quarkonium case, the hybrid potential gets a first contribution already at order $\Lambda^2_{\text{QCD}}/m$.
The corresponding operator does not contribute at LO to matrix elements of quarkonium states as its projection on quark-antiquark color singlet states vanishes.
Hence, spin splittings are remarkably less suppressed in heavy quarkonium hybrids than in heavy quarkonia: this will have a
notable impact on the phenomenology of exotics.}

The coefficients $V_i(r)$ on the right-hand side of Eqs.~(\ref{sdm}) and (\ref{sdm2}) have the form $V_i(r)=V_{oi}(r)+V_i^{np}(r)$, where $V_{oi}(r)$ is the perturbative octet potential and $V_i^{np}(r)$ is the nonperturbative contribution in the case in which $1/r \ll \Lambda_{QCD}$. The octet part can be calculated in perturbation theory while the nonperturbative part is given in terms of purely gluonic correlators 
whose detailed form is given in   \cite{Brambilla:2019jfi} (or they are given in term of generalized Wilson loops with  hybrids as initial and final  states in the case in which $ 1/r \sim \Lambda_{QCD} $).
There are eight of these nonperturbative correlators that have not yet been calculated on the lattice, therefore we fix them on direct lattice determination of the charmonium hybrid masses.
In \cite{Brambilla:2019jfi} we used data from the Hadron Spectrum Collaboration  one set from Ref.~\cite{HadronSpectrum:2012gic} with a pion mass of $m_{\pi}\approx 400$ MeV and a more recent set from Ref.~\cite{Cheung:2016bym} with a pion mass of $m_{\pi}\approx 240$ MeV.  For all the details of the fit we refer to  \cite{Brambilla:2019jfi} .
We take the values $m^{RS}_c(1{\rm GeV})=1.477$~GeV  
and $\alpha_s$ at $4$-loops with three light flavors, $\alpha_s(2.6\textrm{~GeV})=0.26$.

\begin{figure}[!tbp]
 \begin{minipage}[b]{0.5\textwidth}
    \includegraphics[width=\textwidth]{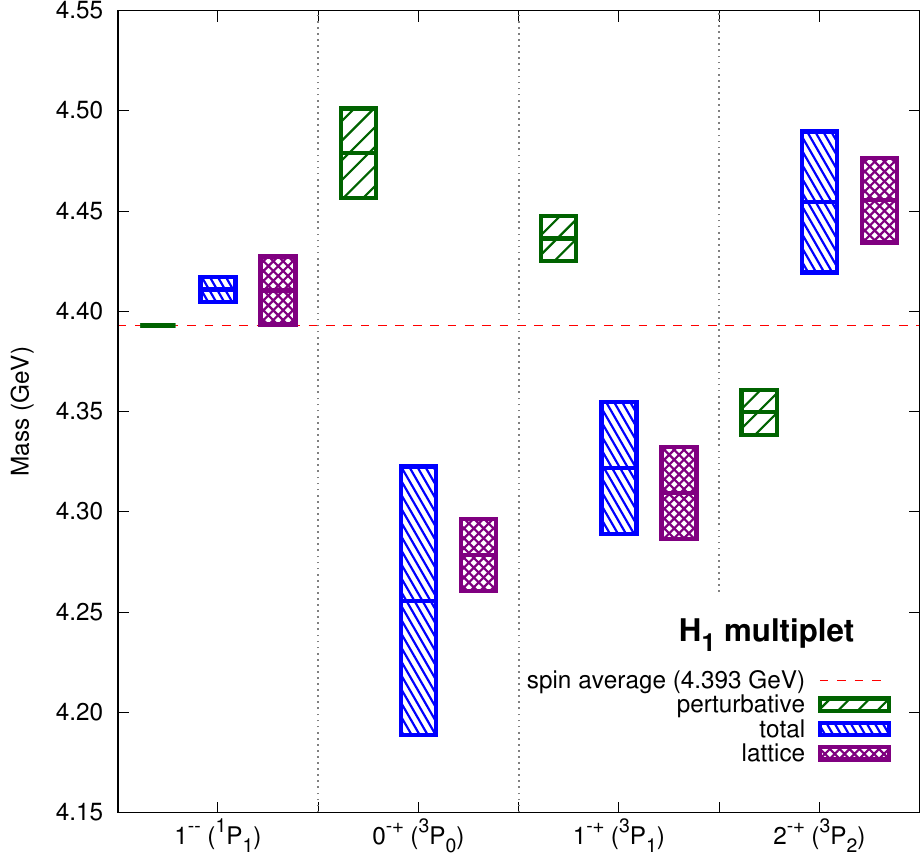}  
     \end{minipage}
\hfill
  \begin{minipage}[b]{0.5\textwidth}
    \includegraphics[width=\textwidth]{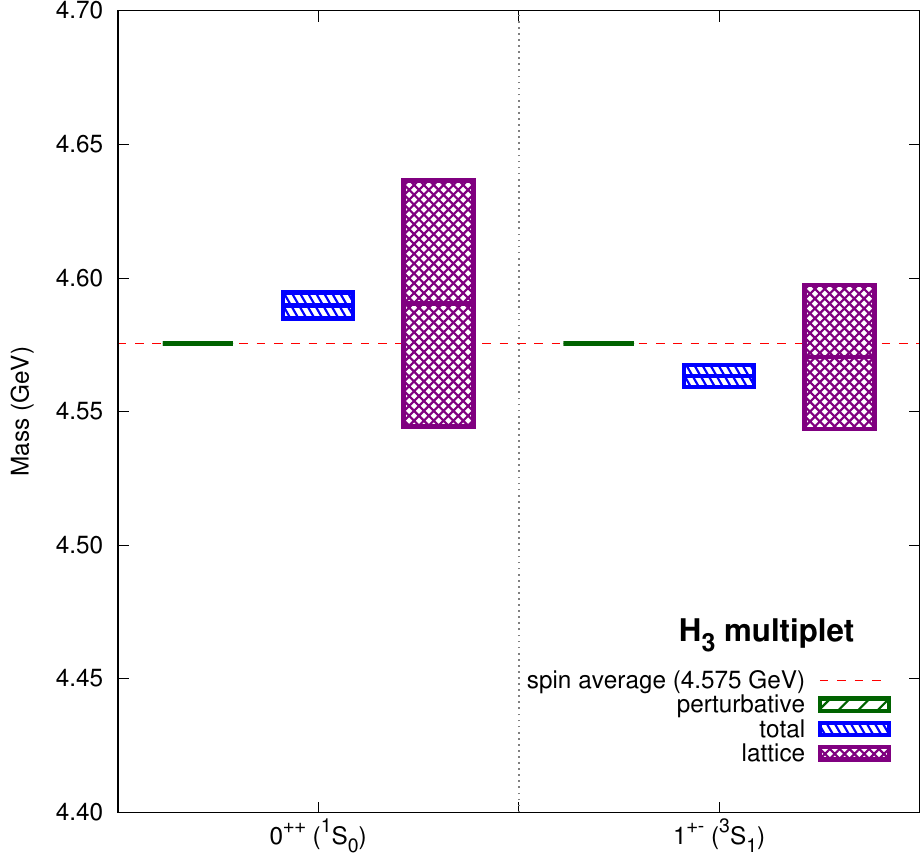} 
 \end{minipage} 
  \caption{Spectrum of $H_1$ and $H3$  charmonium  hybrid multiplets (including the spin effects). 
  The lattice results from ~\cite{Cheung:2016bym}
 [2+1 flavors, $m_\pi = 240$~MeV]    are plotted in purple. In green we plotted the perturbative contributions to the spin-dependent operators in Eq.~\eqref{sdm2} added to the spin average 
 of the lattice results (red dashed line).  In blue we show the full result of the spin-dependent operators of Eqs.~\eqref{sdm}-\eqref{sdm2} including both the perturbative and nonperturbative contributions. The unknown nonperturbative matching coefficients are determined by comparing the charmonium hybrid spectrum obtained from the BOEFT to the lattice data. 
 The height of the boxes indicate the uncertainty as detailed in 
 \cite{Brambilla:2019jfi}, there you find the spectrum for all the multiplets.}
 \label{c}
   \end{figure}

\begin{figure}[!tbp]
\begin{minipage}[b]{0.4\textwidth}
    \includegraphics[width=\textwidth]{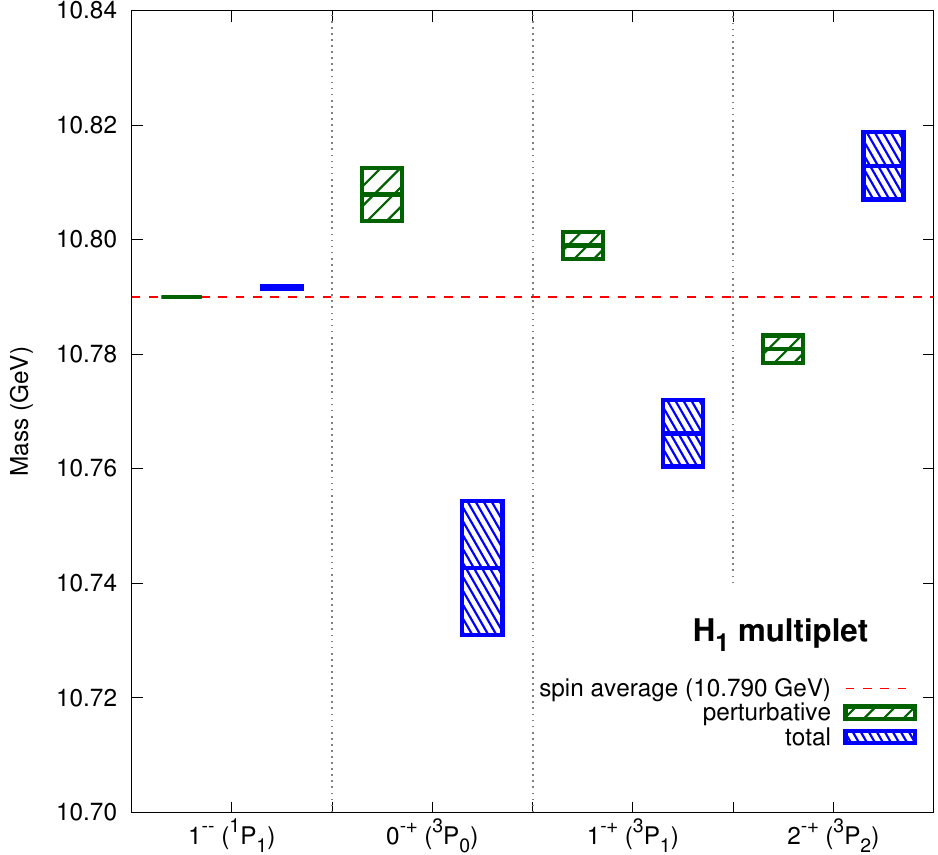}  
     \end{minipage}
\hfill
  \begin{minipage}[b]{0.4\textwidth}
    \includegraphics[width=\textwidth]{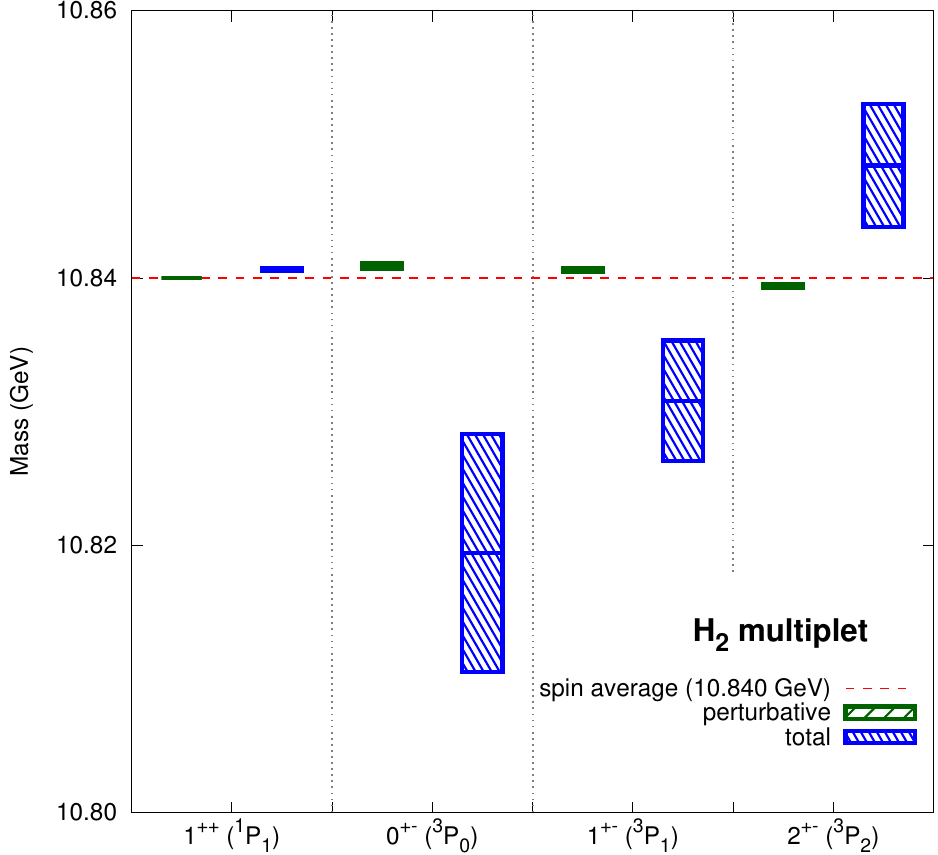} 
 \end{minipage}      
   \caption{Spectrum of $H_2$ and $H_1$   bottomonium hybrids  multiplets,  nonperturbative correlators fixed on charmonium hybrids.
   In green we plotted the perturbative contributions to the spin-dependent operators in Eq.~\eqref{sdm2} added to the 
  spin average of the lattice results (red dashed line).  In blue we show the full result of the spin-dependent operators of Eqs.~\eqref{sdm}-\eqref{sdm2} including both the perturbative and nonperturbative contributions.
   The height of the boxes indicate the uncertainty as detailed in 
 \cite{Brambilla:2019jfi}, there  you find the spectrum for all the multiplets.}
\end{figure}

An interesting feature is that for the spin-triplets, the value of the
perturbative contributions decreases with $J$.  This trend is opposite
to that  of the  lattice results. This  discrepancy can  be reconciled
thanks to the nonperturbative contributions,  in particular due to the
contribution  from $V^{np\,(0)}_{SK}$,  which  is  only suppressed  by
$1/m$,  and has  no perturbative  counterpart.  A  consequence of
this is that a model calculation with a  spin interaction inspired by quarkonium
physics would give the wrong result.

All the dependence on the heavy-quark mass of the $V^{np\,(i)}$ is encoded in the NRQCD matching coefficients $c_F$ and $c_s$.
At leading order in $\alpha_s$ these coefficients are known to be equal to $1$ and the dependence on the heavy-quark mass only appears when the
next-to-leading order is considered. At the order we are working, only the
heavy-quark mass dependence of $c_F$ is relevant, we use its one-loop expression, with the renormalization scale set 
as the heavy-quark mass. Taking this mass dependence into account, we can use the set of nonperturbative parameters (that are flavor independent)
to predict the spin contributions in the bottomonium hybrid 
sector, for which lattice determinations are difficult to obtain. We present the results for two multiplets in Fig. 8, you can find the other multiplets in 
\cite{Brambilla:2019jfi}.

{\it This is a nice example of the huge predictivity that BOEFT plus lattice input can achieve: once the nonperturbative correlators
are fixed on lattice data for charmonium hybrids (or better when these low energy correlators will be directly calculated on the lattice) then we can predict
all the bottomonia (and $B_c$) hybrids multiplets.}

\subsubsection{Tetraquarks, Decays, Production}

BOEFT may be used to describe also tetraquarks.
In this case, it is necessary to compute on the lattice  generalized Wilson loops of the type  given in Eq. (\ref{gamma})
where the operator $P^a_\Gamma$  contains also appropriate light quark operators and besides the $\Lambda^\sigma_\eta$ and 
$\kappa$ also the isospin quantum numbers  $I=0, 1$ have to be considered. The same approach can be used also to describe pentaquarks
\cite{ Brambilla:2017uyf,Soto:2020xpm}. The light quark plays  a similar role to the gluonic excitations in the hybrid case.
These static energies, for which few pioneering lattice studies exist \cite{Sadl:2021bme,Prelovsek:2019ywc,Bicudo:2015kna},
are  a crucial ingredient
to provide for the first time a dynamical description of tetraquarks in QCD.
When these objects will be made available, it will be interesting 
to study the behaviour of the static energies near the avoided 
level crossing with the heavy-light meson-meson thresholds \cite{Bruschini:2021fuw}.
Decays of hybrids and tetraquarks as well as  and mixing of hybrids and tetraquarks
with other quarkonium states could be studied and computed in BOEFT, however
lattice input on few correlators purely depending on the glue would be needed \cite{Oncala:2017hop}.

\subsection{Tetraquarks, molecules,  hadroquarkonium ?}
In this BOEFT/Lattice picture we will have all of the above descriptions contributing \cite{TarrusCastella:2019rit}.
When looking at the actual  
plot of a static energy as a function of $r$ for a state with $Q\bar{Q} q \bar{q} $ or $Q\bar{Q}  g $ we will have different regions: for 
short distance a hadroquarkonium picture would emerge, then a tetraquark (or hybrid) one and when passing the the heavy-light   mesons 
line,  threshold effects  should have to be taken into account and a molecular  picture would emerge.
However QCD would dictate, throught the lattice correlators and the BOEFT characteristics and power counting, which structure would dominate 
and in which precise way.

\subsection{Input needed from the lattice}

Summarizing,  to address  the X Y Z states we need input from the lattice:

\begin{itemize}

\item{} Besides the hybrids static energies we need all the tetraquark static energies in the short and long distance,
extrapolated to continuum, calculated with the generalized Wilson loop mentioned in 3.2. Some pioneering 
calculations exist   \cite{Sadl:2021bme,Prelovsek:2019ywc,Bicudo:2015kna} but more should be done.

\item{} We need lattice calculations  of the nonperturbative correlators entering the spin dependent hybrids and tetraquark potentials: 
they are gauge invariant correlators of chromolectric and chromomagnetic fields and  gluelump operators (without the color octet quark part).

\item{} We need lattice calculations of three point functions containing the information of the quarkonium/hybrids quarkonium/tetraquarks
  mixing.

\item{} Lattice results about the crosstalk of the static energies with a pair of 
heavy-light mesons in the lattice  appeared   \cite{Bali:2005fu,Bulava:2019iut,Bruschini:2021fuw}
 but further investigations appear to be necessary.

\end{itemize}
{\it This will be instrumental to tackle the X Y Z states in QCD.}

\section{Outlook}
We have shown that there is a vital interplay between EFTs and lattice for what concerns Exotics.
NREFTs are  vital to define objects  of great phenomenological interest and give a systematic scheme  to calculate physical observables,
as well as an understanding of the underlying degrees of freedom and dynamics.

NREFTs are based on factorization which allows to  encode all the nonperturbative  effects in low energy correlators, depending 
only on  chromoelectric and chromomagnetic fields. This enhances the predictivity of the theory since the same correlators 
can be used in the charmonium and bottomonium sectors  eliminating for example the difficulties of direct  lattice
calculations of exotics at  the high mass of the bottom.
Moreover, once the lattice has provided the nonperturbative input on the correlators, the calculation of the physical  properties 
ranging from masses, decays, transitions and production, follow inside the BOEFT, which is much simpler and under control 
that repeating a lattice calculation for each of these observables.
Of course there are challenges: the  calculation of  correlator of  chromoelectric and chromagnetic fields display 
bad convergence properties on the lattice see e.g. \cite{Brambilla:2021wqs}    and references therein.
The gradient flow method seems however to help tremendously here, e.g. see \cite{Leino:2021bpz,Leino:2021-157/21}
There may be still difficulties related to change of  renormalization schemes between the continuum of the BOEFT and the lattice 
regularization. For physical objects like the force there are no issues and a gradient flow calculation in continuum can help 
the lattice extrapolations \cite{Brambilla:2021egm}.

We regard the combination of BOEFTs and Lattice QCD as  very promising to attack the physics of the X, Y, Z  on the basis of QCD.
We have seen that Lattice is crucial to gain predictivity in these NREFTs calculations,
on the other hand the NREFTs allow to use lattice  results in an array of problems that often are not  
in its reach  like  nonequilibrium evolution,  production processes and  in general to develop a convenient  and complementary  avenue 
to address exotics.

\end{document}